\documentclass[sigconf]{acmart}

\AtBeginDocument{%
  \providecommand\BibTeX{{%
    \normalfont B\kern-0.5em{\scshape i\kern-0.25em b}\kern-0.8em\TeX}}}

\acmConference[EASE 2020]{Evaluation and Assessment in Software Engineering}{April 15--17, 2020}{Trondheim, Norway}
\acmPrice{XXX}
\acmDOI{}
\acmISBN{}

\setcopyright{none}

\usepackage{adjustbox}

\usepackage{float}
\usepackage{array}
\usepackage{subcaption}
\usepackage{rotating}
\usepackage{xspace}
\usepackage[normalem]{ulem}

\usepackage{enumitem} 

\newboolean{isblinded}
\setboolean{isblinded}{true}
\ifthenelse{\boolean{isblinded}}
{\newcommand\blind[1]{BLINDED\xspace}}
{\newcommand\blind[1]{#1\xspace}}

\usepackage{ifthen}
\usepackage{xcolor}

\newboolean{hide}
\setboolean{hide}{false} 
\ifthenelse{\boolean{hide}}
{
  \newcommand\hide[1]{\nbc{HIDE}{#1}{gray}}
}{
  \newcommand{\hide}[1]{}
}

\newboolean{showfw}
\setboolean{showfw}{false} 
\ifthenelse{\boolean{showfw}}
{
  \newcommand\fw[1]{\nbc{FUTURE WORK}{#1}{gray}}
}{
  \newcommand{\fw}[1]{}
}

\newboolean{showedits}
\setboolean{showedits}{true} 
\ifthenelse{\boolean{showedits}}
{
	\newcommand{\del}[1]{\textcolor{red}{\sout{#1}}} 
	\newcommand{\nbe}[3]{
		{\colorbox{#3}{\bfseries\sffamily\scriptsize\textcolor{white}{#1}}}
		{\textcolor{#3}{\sf\small$\blacktriangleright$\textit{#2}$\blacktriangleleft$}}}
}{
	\newcommand{\del}[1]{} 
	
	\newcommand{\nbe}[3]{}
}

\newboolean{showcomments}
\setboolean{showcomments}{true} 
\newcommand{\id}[1]{$-$Id: scgPaper.tex 32478 2010-04-29 09:11:32Z oscar $-$}

\ifthenelse{\boolean{showcomments}}
 {
 	\newcommand{\nbc}[3]{
 		{\colorbox{#3}{\bfseries\sffamily\scriptsize\textcolor{white}{#1}}}
		{\textcolor{#3}{\sf\small$\blacktriangleright$\textit{#2}$\blacktriangleleft$}}}
	
 }{
 	\newcommand{\nbc}[3]{}
 	
 }

\usepackage[most]{tcolorbox}
\ifthenelse{\boolean{showedits}}
{
  \newtcolorbox{inserted}{%
       title=Inserted text:,
       colframe=blue,colback=blue!5!white,
       breakable,
       leftrule=0mm, 
       bottomrule=0mm,
       rightrule=0mm,
       toprule=0mm,
       arc=0mm, outer arc=0mm,
       oversize
  }
  \newtcolorbox{deleted}{%
       title=Deleted text:,
       colframe=red,colback=red!5!white,
       breakable,
       leftrule=0mm, 
       bottomrule=0mm,
       rightrule=0mm,
       toprule=0mm,
       arc=0mm, outer arc=0mm,
       oversize
  }
  \newtcolorbox{refactored}{%
       title=Rewritten text:,
       colframe=blue,colback=red!5!white,
       breakable,
       leftrule=0mm, 
       bottomrule=0mm,
       rightrule=0mm,
       toprule=0mm,
       arc=0mm, outer arc=0mm,
       oversize
  }
}{

}

\newcommand{\commented}[1]{}

\newcommand{\eg}{\emph{e.g.,}\xspace}
\newcommand{\ie}{\emph{i.e.,}\xspace}
\newcommand{\etal}{\emph{et al.}\xspace}
\newcommand{\etc}{\emph{etc.}\xspace}

\makeatletter                  
\def\mdseries@tt{m}      
\makeatother                   

\begin{document}

\title{Caveats in Eliciting Mobile App Requirements}


\author{Nitish Patkar}
\orcid{1234-5678-9012}
\affiliation{%
 \institution{University of Bern}
}
\email{nitish.patkar@inf.unibe.ch}

\author{Mohammad Ghafari}
\orcid{0000-0002-1986-9668}
\affiliation{%
 \institution{University of Bern}
}
\email{mohammad.ghafari@inf.unibe.ch}

\author{Oscar Nierstrasz}
\orcid{0000-0002-9975-9791}
\affiliation{%
 \institution{University of Bern}
}
\email{oscar.nierstrasz@inf.unibe.ch}

\author{Sofija Hotomski}
\affiliation{%
 \institution{University of Zurich}
}
\orcid{}

\renewcommand{\shortauthors}{Patkar, et al.}

\begin{abstract}
Factors such as app stores or platform choices heavily affect functional and non-functional mobile app requirements.
We surveyed 45 companies and interviewed ten experts to explore how factors that impact mobile app requirements are understood by requirements engineers in the mobile app industry.

We observed the lack of knowledge in several areas. 
For instance, we observed that all practitioners were aware of data privacy concerns, however, they did not know that certain third-party libraries, usage aggregators, or advertising libraries also occasionally leak sensitive user data.
Similarly, certain functional requirements may not be implementable in the absence of a third-party library that is either banned from an app store for policy violations or lacks features, for instance, missing desired features in ARKit library for iOS made practitioners turn to Android.
  
We conclude that requirements engineers should have adequate technical experience with mobile app development as well as sufficient knowledge in areas such as privacy, security and law, in order to make informed decisions during requirements elicitation. 

\end{abstract}



\maketitle

%
%

\section{Introduction}
\label{sec:introduction}
The number of available mobile apps is constantly increasing ever since app hosting platforms came into existence ten years ago.
For example, the Apple App Store now hosts about 2.2 million mobile apps.
About 28\% of the installed apps are however uninstalled within 30 days, showing that such apps failed to satisfy users.

\newpage

Although there are several similarities between mobile and desktop apps, such as the availability of app stores and multiple available platforms, research has also shown several differences between them~\cite{Mine13,Wass10}.
Additionally, the availability and possibility of interpreting data obtained from various onboard sensors on mobile phones allow apps to know much more about their users than the users might willingly wish to share~\cite{Zang15,Xu15,Chri16}, thus jeopardizing the privacy of millions of users. 
This study does not assume that requirements elicitation process is different for mobile apps, instead we simply reflect on the factors that are more prevalent in the mobile domain, and investigate how they are perceived in industry. 

We conducted an industry survey in which 45 mobile app development companies from Switzerland, Germany, and the Czech Republic participated.
In addition, we interviewed ten industry experts from Europe and India working in the mobile app industry and responsible for requirements elicitation. 
Consequently, we aim to answer the research question: \emph{``Which factors are essential for mobile app requirements, and how well are they understood by practitioners?''}

Most participants who claimed to be responsible for requirements elicitation in fact shared roles such as CEO, business analyst, or marketing head, and lacked knowledge in several areas.
For instance, most of them had no direct technical experience with mobile app development. 
Consequently, they fail to understand how, for instance, non-functional requirements affected by usability and performance, are also compromised by third-party libraries including app analytics that have been reported to leak sensitive user data. 
In several cases, we observed that the practitioners lacked additional knowledge of app security, privacy and law.
Consequently, they relied heavily on developers' knowledge of security and on external lawyers for privacy matters, failing to communicate the implications of certain requirements choices to their clients. 
We believe that discussing the implications of the observed factors \ie platform choices, app store policies, third-party libraries, NFRs, and app type for the mobile app requirements will help RE practitioners to make informed decisions.

The remainder of the paper is structured as follows:
In \autoref{sec:methodology}, we explain the protocol we followed for the survey and the interview.
We then present and discuss our discoveries in \autoref{sec:discoveries}.
Related work is discussed in \autoref{sec:related-work}, and finally \autoref{sec:conclusion} concludes our findings.

\newpage
%
%

\section{Research method}
\label{sec:methodology}
To answer the research question we conducted a qualitative exploratory study and followed the procedure described by Patton \etal~\cite{Patt90}.
We first performed a survey that ultimately led to the interviews to obtain more detailed information from industry representatives.
The survey and interview instruments, and supporting graphs are available online.\footnote{\label{supplement-url}\url{https://figshare.com/s/87a07b7c47f37952eef8}}

\subsection{Survey}
\label{sub:study-design-survey}
The online survey was performed from October 2018 to late January 2019.
The invitations for participation were sent manually to 276 mobile app companies; it was also advertised on \emph{LinkedIn} and \emph{Twitter}.
We specifically looked for people who claimed to be responsible for requirements elicitation in mobile app companies.

\emph{Preparation.} 
The survey questions were selected based on the themes evident from a literature review of 60 publications where the authors identified 24 elicitation techniques that target mobile apps, out of which only half were empirically evaluated~\cite{Patk19}.
The questions in the survey have been validated and improved over several iterations to accurately answer the proposed research question.
It consists of five sections:
The first section characterizes the participants, while the remaining four sections focus on various aspects of mobile app development, such as the \emph{application domain}, \emph{elicitation parameters}, \emph{communication}, and the \emph{developers}.

\emph{Participants.} 
In total 45 people with various roles, such as requirements engineer, developer, or even CEO, participated in the survey. 
They all claimed to be responsible for the requirements elicitation in their respective companies.
About 60\% of them (\ie 27) had more than five years of experience in this area.
About ten participants shared multiple roles within their company.

\emph{Companies, products, and services.} 
We received responses from both product-based and service-oriented companies.
Their apps mainly belonged to but were not limited to categories such as travel, transport, logistics/marketing, business, and productivity.

\subsection{Interview}
\label{sub:study-design-interview}
We designed an interview with 52 questions to thoroughly investigate the commonly emerging themes from the survey responses.

\emph{Preparation.}
The questions in the interview have been validated and improved in two major iterations:
First, they were reviewed and discussed by the collaborators of this paper, and second, they were validated in a pilot interview with an external expert in  mobile app development.
The pilot interview is not included in the main study.
The interview consists of four sections:
The first two sections characterize the interviewee and the companies, the third section explores the state of the art requirements elicitation in the mobile app industry, and finally, the fourth section explores the factors that might affect requirements gathering for mobile apps.

\begin{table*}[t]
\centering
\caption{Interviewees, their roles, and details regarding their employers}
\label{tab:participants-overview}
\begin{adjustbox}{width=\textwidth}
    \begin{tabular}{@{}cclccccccc@{}}
    \toprule
    \begin{tabular}[c]{@{}c@{}}Participant\\ ID\end{tabular} & \begin{tabular}[c]{@{}c@{}}Company\\ ID\end{tabular} & Participant Role   & \begin{tabular}[c]{@{}c@{}}Experience in \\ years with RE\end{tabular} & \begin{tabular}[c]{@{}c@{}}Company\\ Type\end{tabular} & \begin{tabular}[c]{@{}c@{}}Number of\\ Published Apps\end{tabular} & \begin{tabular}[c]{@{}c@{}}\% RE budget \\ Spent\end{tabular} & \begin{tabular}[c]{@{}c@{}}Number of\\ Employees\end{tabular} & \begin{tabular}[c]{@{}c@{}}Which Markets\\ They Target\end{tabular} & Country of Origin \\ \midrule
    P1                                                       & C1                                                   & CEO                & 5                                                                      & Service                                                & 15                                                                 & 15-20                                                         & 35                                                            & Switzerland                                                         & Switzerland       \\
    P2                                                       & C2                                                   & Team Lead          & 6                                                                      & Service                                                & 13                                                                 & 10-20                                                         & 29                                                            & Switzerland                                                         & Switzerland       \\
    P3                                                       & C2                                                   & Project manager    & 2.5                                                                    & Service                                                & 13                                                                 & 10-20                                                         & 29                                                            & Switzerland                                                         & Switzerland       \\
    P4                                                       & C3                                                   & Team Lead          & 6                                                                      & Service                                                & 9                                                                  & 5-10                                                          & 7                                                             & Switzerland                                                         & Switzerland       \\
    P5                                                       & C3                                                   & CEO                & 6                                                                      & Service                                                & 9                                                                  & 5-10                                                          & 7                                                             & Switzerland                                                         & Switzerland       \\
    P6                                                       & C4                                                   & Project manager    & 20                                                                     & Product                                                & 1                                                                  & 10                                                            & 8                                                             & Switzerland                                                         & Switzerland       \\
    P7                                                       & C5                                                   & Director           & 7                                                                      & Service                                                & 10                                                                 & 40-50                                                         & 31                                                            & Europe, India, USA                                                  & India             \\
    P8                                                       & C6                                                   & Agile consultant   & 15                                                                     & Service                                                & 10+                                                                & $\sim$20                                                      & NA                                                            & Europe, Asia                                                        & Switzerland       \\
    P9                                                       & C7                                                   & Business developer & 10                                                                     & Service                                                & 20                                                                 & 15-20                                                         & 150                                                           & Switzerland                                                         & Switzerland       \\
    P10                                                      & C8                                                   & Product Owner      & 7                                                                      & Both                                                   & 1                                                                  & $\sim$1                                                       & 100                                                           & Netherlands                                                         & Netherlands       \\ \bottomrule
\end{tabular}
\end{adjustbox}
\end{table*}

\emph{Participants.}
Survey respondents who were interested in participating in an interview were invited by email.
We also invited other practitioners through our contacts, who we know were responsible for requirements elicitation and additionally advertised the interview on \emph{LinkedIn} and \emph{Twitter}.
Subsequently we interviewed ten practitioners who agreed to participate in an in-depth interview.
We observed that in most small companies in our study, the role has very little to do with actual expertise in RE.
People with different roles ranging from business analysts to CEO were responsible for requirements elicitation.
This explains why the implications of requirements elicitation are not well understood in practice.
An overview of the interviewees can be found in \autoref{tab:participants-overview}.
For instance, participant P1 works in company C1 as a CEO.
This Service-based company has developed 15 apps, and it usually spends 15-20\% of its budget on RE.
The company has 35 employees who are based in Switzerland and target the Swiss market.
\noindent
Most of the interviewees had a computer engineering background and they all could speak English fluently.
They had varying degrees of experience with requirements elicitation ranging from 2 to 20 years, and most of them learned the corresponding elicitation techniques through experience at the workplace.
Importantly, a few of them who have an academic background mentioned that they no longer have free access to academic publications, which deprives them of the latest advances in academia.

\emph{Companies, products, and services.}
An overview of the interviewees' companies can also be found in \autoref{tab:participants-overview} as well.
The companies were based mainly in Switzerland.
The companies were of different sizes in terms of number of employees ranging from seven up to 150 employees.
Similarly, the numbers of mobile apps they have delivered varied considerably, ranging from one up to twenty, which shows their varied experience in this field.
Except for one company, all were service-oriented, which indicates a risk associated with developing an innovative app in the absence of financial funding and through market research, especially for small-sized companies.

\emph{Data collection and analysis.}
All ten interviews were carried out between February and March 2019 and were conducted by the first author.
Although P10 preferred to fill in an interview template due to time constraints, other interviews were conducted face-to-face.
The duration of the interviews was between 60 and 120 minutes.
The voices of the interviewees in all face-to-face interviews were recorded, as was the video signal in all video call interviews, with permission from the interviewees.
In order to analyze the data, all interviews were digitally transcribed and coded.
We created an initial list of codes based on the survey results and complemented it with codes that emerged while reviewing interview transcripts.
The codes were used to group related answers and to compare them to each other.

%
%

\section{Discoveries}
\label{sec:discoveries}
This section discusses the factors that are known to the research community, but that are not reviewed in prior work from the requirements perspective.
In particular, we discuss how (i) platform and third-party libraries, (ii) app stores and their policies, (iii) RE practitioner's experience and expertise regarding NFRs, (iv) learning from app usage preferences, and (v) app type are affecting app requirements, and how practitioners perceive these factors for the success of the app.
These factors emerged essentially from the survey responses, and were later discussed in depth during the interviews.
For each factor, we report our own empirical observations from the survey and the interviews, provide a discussion referring to the existing literature wherever relevant, finally followed by conclusions.
The summary of all the important findings can be seen in~\autoref{tab:main-findings}.

\subsection{Platform and third-party libraries}

Platform selection depends on several obvious factors such as cost, budget, and market research. 
Besides these factors, a few other non-obvious factors such as the role of app stores, or third-party libraries are equally influential, and can directly impact functional requirements.
These factors are inter-related and inter-dependant, and their relevance to app requirements should be made explicit.

\subsubsection{Empirical observations}
\emph{Survey.} 
According to the survey participants (\eg total nine), \emph{only few} customers have a clear idea about their choice of platform, and in spite of the limited funds, without any exception they would like to be present ``everywhere.''

\emph{Interview.}
P7 mentioned that apart from platform features, customer bases in different geographical markets also influence the platform selection decision.
According to him, Indian customers prefer Android, whereas European and U.S. customers prefer iOS as their first platform for app releases.

One of the experts (\ie P8) further said that they usually choose the platform that best supports the customer's requirements, which means that the same app designed for different platforms might use different features to \emph{fully utilize the underlying platform capabilities}.
For instance, company C3 chose the Android platform for one of their augmented reality projects as the bundled library \emph{ARCore} provided features unavailable in iOS' \emph{ARKit}.
All the interviewees further agreed that when adopting the Android platform, additional properties such as the minimum OS version supported or numerous screen sizes must also be considered to optimize an app.

Interestingly, various challenges posed by platform ecosystems have led interviewees to consider \emph{hybrid apps}\footnote{A hybrid app is developed using browser supported language \eg HTML, CSS, JavaScript that runs only in a browser shell and has access to the native platform layer.} over native Android and iOS apps. 
According to P1, in most cases customers choose hybrid apps over native ones mainly for economic reasons because they are cheaper to build as they do not require dedicated iOS or Android developers.
Interviewees, however, had different opinions about hybrid apps in the mobile app market.
P5, for instance said that hybrid platforms add ``yet another dependency'' in addition to native layer and hence, he fears that developers tend to lose control over security.

\subsubsection{Discussion}
One should select a mobile platform that brings justice to the end-user requirements.
We identified three key themes from our observations:
\begin{itemize}
    \item Platform selection depends on several factors such as cost, budget, market research and platform ecosystem.
    \item End users, customers and developers can have different concerns regarding platform selection.
    \item Functional requirements are affected while selecting a particular platform, and the opposite is true as well, sometimes resulting in favouring hybrid apps.
\end{itemize}

From the customer's point of view, platform selection is based on market research and the number of potential users, \eg in Switzerland iOS is prevalent, while in India it is Android. 
Customers also have stringent budget concerns.
However, they may not be experts when a decision must be made.
Kumar \etal have shown how several mobile app projects suffer from low funding, eventually resulting in low-quality apps~\cite{Kuma16}.
They mention that it is primarily because mobile app development is perceived to be cheap by customers, which in fact, is incorrect.  

On the other hand, app developers and RE practitioners seem to be more concerned about device fragmentation as it leads to increased cost and efforts.
The aforementioned issues have been discussed in previous publications, notably stressing the direct impact on the initial development and post-release maintenance cost~\cite{Chri11,Trac12}.
The effect of fragmentation on testing and eventually on platform selection is also evident from literature. 
Miranda \etal report that the variety of compatible devices and OS versions for Android makes Android app testing much easier than iOS apps~\cite{Mira14}.
Furthermore, they also point out that Java as a programming language is more widespread than Objective-C, both at universities and private companies, making it easier to find Android developers.
To the best of our knowledge, there is no evidence of potential solutions in the literature to the conflicting concerns regarding platform selection.

Beyond these commonly known factors, the platform ecosystem, including app stores and third-party libraries, affect app requirements and contribute to the platform selection decision.
Certain functional requirements may not be implementable on a specific platform due to, for instance, a missing third-party library.
Both app stores and third-party libraries impose additional challenges to requirements elicitation.
Third-party libraries are heavily used in developing mobile apps, for instance, Minelli \etal from their analysis of a corpus of Android apps found that external calls represented more than 75\% of the total number of method invocations~\cite{Mine13}.
The recent work by Derr \etal discusses an imminent risk of misuse by third-party libraries in Android as they inherit the access rights of their host apps~\cite{Derr18}.
Similarly, Grace \etal have reported how advertising libraries used by developers were collecting user's private data; a few even fetch and run code from the internet~\cite{Grac12}.
Only during development when developers discover that a certain third-party library that is vital in fulfilling a functional requirement is unavailable, or is banned from the app store for policy violations, requirements have to be renegotiated with their clients.
We discuss further implications of app stores and third-party libraries on requirements in~\autoref{subsec:app-stores} and ~\autoref{subsec:app-analytics} respectively.  

Fragmentation and app store related issues have made practitioners consider hybrid frameworks instead of a specific native platform although with some skepticism.
The role and potential of hybrid apps has been discussed in great detail in the literature, and we also explore this later in~\autoref{subsec:app-type}. 

In conclusion, we found that all practitioners were aware about platform fragmentation and the involved cost, however, they failed to provide deeper insights into issues that might arise due to choosing compromised third-party libraries, such as data leaks.
Although, all of them were aware of hybrid frameworks, most of them were skeptical about switching from native to hybrid apps even in cases where it would make sense.
RE practitioners have to be aware about technical alternatives and their feasibility to be able to proactively discuss the shortcomings of the requirements up front. 
Such discussions and decisions can save app companies a lot of time and efforts, and money for customers.

\subsection{App stores and their policies}
\label{subsec:app-stores}

App stores are important for both end users and developers as they simplify the app discovery and distribution process.
Although the importance of the app stores for app developers has been discussed extensively, their impact on requirements is ill understood.

\subsubsection{Empirical observations}
\emph{Survey.} 
There were no dedicated questions on the role of app stores in our survey.
The theme emerged from the survey responses and it was subsequently added to the interview instrument for further discussion with practitioners.

\emph{Interview.} 
App stores and their \emph{time-consuming review processes} have affected the company's workflow in the past, said P5, so they had to be careful while accepting customer requirements to avoid any potential conflicts.
Especially hot fixes that must be released immediately present a severe problem, as all the interviewees agreed.
The complex app release processes introduce additional work for the development teams and the requirements engineers.
They must properly plan their release cycles and consider a tradeoff between fast deployment and low costs, or as C3 puts it: ``more releases mean more money as they demand more time.''

Sudden guideline and \emph{policy changes of app stores} also become problematic.
Interestingly, P4 said that their contracts clearly state that app rejection from app stores shall not be their responsibility.
Concerning technical aspects, companies such as C3 have faced app store restrictions like the maximum allowed number of methods per app, or the maximum allowed app size.
Hence, they nowadays reconsider every requirement about the inclusion of external libraries and media content.
Nevertheless, P1 and P4 mentioned that the current state is improving with newer policies.

\subsubsection{Discussion}
Although app stores exist for desktop apps, they play a far more important role in the case of mobile apps due to the huge volume of downloaded apps.
When Nokia still held the major app market share, Holzer \etal had predicted that mobile app marketplaces from Google and Apple were going to change the app development trends, and consequently, they indicated that developers must familiarize themselves with platform strategies~\cite{Holz11}.
Ten years since then, we see an explosion of mobile apps, and the app market is by far dominated by only two major platforms \ie Android and iOS.
We drew three conclusions from our observations:
\begin{itemize}
    \item Customers have no concerns regarding app stores, but they are important for both end users and developers as they simplify the app distribution and discovery process
    \item Different app stores have different business models and policies
    \item Policy changes by app stores are affecting app development companies by causing unintentional rework or in worse cases app denial.
\end{itemize}

App stores are usually not a concern for customers, but they play an important role for end users and developers for different reasons.
For end users, it is a one-stop marketplace to search for their desired apps, enabling easy download and installation, and additionally providing them with useful data such as user ratings and comments to make informed decisions.
User reviews help developers as well to correct their apps by identifying new requirements.
User review analysis has gained much traction in academia, especially in the RE community as later discussed in~\autoref{subsec:app-analytics}. 

For developers, app stores matter in several ways:
(i) they allow developers to distribute their app easily and increase their app's discoverability, which is also evident from a study of questions posted on StackOverflow~\cite{Rose16};
(ii) different app stores charge differently;
(iii) different app stores have different policies regarding app design, usage of third-party libraries \etc; and
(iv) policies change, so developers have to keep themselves informed and align their workflows accordingly.
Some of these concerns are reflected in the literature.

The difference in business models is highlighted by Holzar \etal as they point out that besides distribution, app stores take care of billing and advertising for a certain commission (usually about 30\%), which helps developers to increase the visibility of their apps on the platform~\cite{Holz11}.
However, the cost of publishing and distributing apps through app stores also differs significantly for Android and iOS platforms.
It costs developers a one-time payment of about 25\$ for Google Play, whereas the Apple App Store charges annually, which can range anywhere from 99\$ to 299\$~\cite{Mira14}.

To the best of our knowledge very few studies discuss the role of app store requirements and policies on the app requirements and workflow. 
Notably, in their study, Joorabchi \etal highlight the need for testing APIs by app stores so that app developers can check their code for any guideline violations~\cite{Joor13}.
Similarly, Holzar \etal studied the app distribution process and subsequently outlined the implications of app store centralization.
In particular, they claim that such centralization limits the freedom of developers.
For instance, Apple's App Store decides which apps will be promoted over others~\cite{Holz11}.
In a recent exploratory study Al Subaihin \etal conducted interviews with mobile development team managers followed by a questionnaire to gather qualitative data on app engineering practices.
They report that 54\% of their survey respondents claimed to adopt a release strategy that is influenced by the app store's regulations \ie by review and approval period~\cite{Alsu19}.
Shortly after Google introduced revised policy changes to maintain the quality of the published apps in 2015, several apps have been removed from Google Play.
The effects of policy changes are reflected in literature as well.
Wang \etal studied 790K removed apps from Google Play to explore and understand the factors responsible for policy violations~\cite{Wang18}.
For instance, they discovered that the apps targeting children should never violate the COPPA policy.\footnote{http://www.coppa.org/}
Interestingly, most of the removed apps were published by \emph{spamming developers}.
Among the important reasons they identified for app removal were apps being classified as either malicious, privacy-risk, spamming, or fake by Google.
The exact parameters that Google uses to flag an app as \emph{malicious} are unclear.
The parameters for flagging \emph{spamming} apps are more explicit, such as those using excessive or inappropriate metadata, especially with misleading references to other apps or products.

In conclusion, we observed that all practitioners were aware of the impact app store policies have on their development workflow.
However, they never discussed the problematic app store policies with their clients during requirements negotiation.
Hence, it has become necessary not just for app developers, but also for RE practitioners, to fully understand what different app stores offer them, and align their development, publishing, and marketing strategies with app store policies.
Customers usually aren't aware of involved complications.
Hence, RE practitioners need to be aware of the latest policies (on top of ecosystem offerings such as third-party libraries) from different app stores to be able to discuss those with customers.
Certain functional requirements may not be implementable because the third-party library is banned from a specific App Store for policy violations.
Such issues must be raised with the customer up front, otherwise, they may lead to considerable rework for developers, and in the worst case renegotiation.


\subsection{Non-functional requirements}
\label{subsec:experience}
Non-functional requirements are important in general, but for mobile apps they are even more vital as they directly influence end-user acceptance.
The General Data Protection Regulation (GDPR)\footnote{https://eugdpr.org/} has fostered awareness regarding data privacy, however, eliciting other non-functional requirements is still ill-supported and can only benefit from practitioner expertise.

\subsubsection{Empirical observations}
\emph{Survey.}
There were no dedicated questions on the role of app stores in our survey.
This theme emerged from the survey responses and it was subsequently added to the interview for further discussion with practitioners.

\emph{Interview.}
All the interviewees mentioned that the \emph{GDPR improved the privacy awareness} of customers who are now becoming sensitive to data privacy requirements.
Most app developers maintain a GDPR-related checklist that they consult during app development.
Other non-functional requirements like security, however, are only considered for basic measures such as maintaining secure data flows during communication.
Except for C2 and C6, which are rather big companies, all companies outsource the legal work to the experts, \eg preparation of contracts and data policy assistance.

Sometimes inter-personal skills alone are not enough to elicit certain requirements, particularly non-functional ones.
Eliciting such requirements requires experience.
For instance, experience is needed to elicit requirements for unpredictable situations such as handling network drops in tunnels to make app usage a pleasant experience.
Reflecting on this, P1 told us of an experience: 
An enterprise app they built for a company suddenly stopped working for a few employees as it could no longer connect to the server, since those employees were connected through a hotel's wireless network that was protected by a captive gate.\footnote{A captive gate forwards all requests to the login page instead of the application server until the user has been authenticated.}
He said such situations are extremely hard to foresee and can only be tackled through experience.
Similarly, P9 mentioned that for mobile apps, users need to manually install the updates, a process that the developers do not control.
This can cause the back end to get out of sync if not handled properly.

\subsubsection{Discussion}
Customers are not good at dictating NFRs, RE practitioners must formulate those themselves.
We identified three recurring themes from our observations:
\begin{itemize}
    \item eliciting NFRs requires experience and expertise;
    \item in the case of mobile apps, NFRs such as data privacy can be unintentionally compromised by selecting a certain third-party library; and
    \item additionally, device capabilities are an important factor for app's performance and acceptance.
\end{itemize}

\noindent
RE practitioners require multifaceted expertise to elicit NFRs:
(i) \emph{Awareness about security and data privacy risks}.
People we interviewed had no idea about security whatsoever; they relied heavily on the developer's knowledge, and developers in turn relied on platform documentation.
Recent research has shown that security issues are prevalent in mobile apps, jeopardizing the privacy and security of millions of users worldwide~\cite{Ghaf17, Gadi18}.
Interestingly, they discovered that old IDE versions had limited support for identifying security violations resulting in countless security issues to be missed by the developers. 
Similarly, Jain \etal discuss the complications of using personal mobile phones in a corporate environment; specifically, as it poses a challenge to enforce corporate policies causing an intentional or unintentional corporate data leak.
They also briefly discuss several security risks for mobile apps such as using custom cryptographic algorithms instead of standard algorithms that can compromise data confidentiality, or failing to disable insecure OS features in mobile apps can result in sensitive data ending up in the web caches or global OS logs~\cite{Jain12}.

Privacy has become a serious concern for many. 
We found a lack of awareness among practitioners about potential factors that can compromise user privacy.
For example, the use of third-party libraries and any unintentional use of trackers can compromise the security and data privacy of users.
Third-party trackers are commonly used by app developers to gather user information, which helps them eventually to build a detailed user profile.
Such a user profile is helpful to draw inferences about shopping habits, socioeconomic class, or political opinions \etc of the user.
Large data aggregators who operate as data brokers (accumulating and selling user information) such as Acxiom and BlueKai collect data from a variety of sources, in most cases implementing user tracking technologies such as cookies on the host device with the consent of app developers~\cite{Mont15}.
End users are often unaware of such data leaks.
\newline
\noindent
(ii) \emph{Awareness about legal complications}.
Companies will need a much deeper understanding of problems arising due to diverse legal regulations across the world when the apps must be released globally.
Not only that, when apps are using third-party libraries or usage trackers that fall outside their jurisdiction, they need to understand the consequences.
Although outsourcing legal work might be affordable for big companies, how a large number of individual app developers and small companies will cope with this is an open question.
Third-party trackers mentioned earlier have been revealed to be a highly transnational problem as many of them are operating outside the jurisdiction of the EU.
The study by Binns \etal outlines some significant legal compliance challenges such as cross-jurisdictional data flow, profiling, and rights and obligations regarding children, arising from the use of trackers~\cite{Binn18}.
\newline
\noindent
(iii) \emph{Awareness about mobile device capabilities}.
Device fragmentation affects performance and perceived app quality.
End users who eventually accept or reject mobile apps place high value on the quality of the apps.
App quality mainly depends on the device capabilities and the app's usability.
Different devices mean an app might behave differently for different users; certain features may not be available for those with older OS versions or the apps might run slower.
Noei \etal in their exploratory study found that the perception of app quality strongly relates to powerful CPUs and other device attributes, and it is not limited to app attributes such as code size~\cite{Noei17}.
Similarly, it is also hard to quantify the implications of security measures on user experience. 
For example, how much extra power will be needed for the added encryption and decryption on variety of devices while choosing HTTPS over HTTP is still unknown~\cite{Thom14}.
Usability and issues with usability testing of mobile apps have been studied in several publications~\cite{Zhan05, Hoeh15}.

In conclusion, we observed that all practitioners were aware of data privacy concerns, however, they did not know that certain third-party libraries, usage aggregators, or advertising libraries also occasionally leak sensitive user data.
They all agreed that app security is a major concern, and that there is a lack of security experts in the industry, especially in many small mobile app companies.
Requirements must be elicited by experts in the field, and, therefore, the app development companies tend to outsource their legal work to legal experts.
Similarly, requirements engineers will proactively need to consult the latest literature and experts in the area of security and privacy to raise awareness about identified breaches in existing tools or libraries.


\subsection{Learning about the end users}
\label{subsec:app-analytics}

Market research is important for RE in general, however RE practitioners usually have little to do with it.
Hence, app development companies rely heavily on other means such as app analytics to learn more about their users.

\subsubsection{Empirical observations}
\emph{Survey.}
All but four of the participants said that end users are anonymous to them.
The locality of the end users, however, appears to be important: 29 said it matters.
The ethnicity of the end users, on the other hand, is not a concern for many, \ie 31 said it has never mattered before.
About 33\% of the participants said that they consider only young people and adults as their end users.
Most of the participants (34, \ie 75\%) agreed that user feedback posted on app stores matters to their customers, but they process it almost without exception manually, as it is usually of low volume.
In our survey 18 participants (\ie 40\%) said they never evaluate user feedback on social media platforms such as Twitter or Facebook.

\emph{Interview.}
When asked about their end users, all interviewees said that the end users were always anonymous; the reason we believe is most of the corresponding companies were service-oriented.
They also mentioned that their customers often did not have any access to the end users either.
Worse, market research was entirely the customer's responsibility.
According to all of them, when apps are released globally it is hard for requirements engineers to make assumptions about end user preferences.
Ethnicity or locality were not concerns for any of them.

Acceptance testing is vital to uncover missing requirements and improve usability.
P9 told us that it is hard to get real end users for beta testing, and they end up testing the app with the people they know, which were mostly their own colleagues.
This poses a great risk of product failure.
All interviewees agreed that it is common practice to start making assumptions about the users whenever market data is missing.

Another way to eliciting anonymous end user preferences is through app analytics.
Post-release app usage statistics, \eg provided by \emph{Google Analytics}, are a very powerful tool to improve existing requirements.
Interestingly, although none of the interviewees carry out pre-release user and market research themselves,
all of them stated that post release analytics services can support app developers in decision making of the global audience's preferences.
Interviewees P2, P3, and P5 said that there is usually no extra budget planned for app analytics, instead another (additional) contract is usually set up.

New requirements can also be discovered through app reviews.
Importantly, similar to the survey participants, all the interviewees were hardly aware of research in the area of opinion mining.

\subsubsection{Discussion}
There are several ways to learn about end-user requirements, and each has its own complications.
We identified four important themes from our observations:
\begin{itemize}
    \item market research is the client's responsibility, but it is rarely reliable;
    \item developers use other means to learn about the user requirements such as through beta testing, trackers, analytics, and app reviews;
    \item beta testing suffers from a lack of access to the actual users; trackers and analytics have privacy breaches; and
    \item app review analysis is more popular in academia than in the industry.
\end{itemize}
\noindent
We see that app development companies rely on their customers to get insights about the potential end users.
The quality of market research depends heavily on the skills and the reach of the customers.
We learned that customers rarely have access to enough end users for beta testing.
We suspect that they make far too many assumptions, specifically because their app users were located in the same country.
Such assumptions fail if the app is to be released globally.
The issues with beta-testers are previously reported by Joorabchi \etal in 2013, where they mention that \emph{``beta-testers are in the order of dozens and not thousands''}~\cite{Joor13}.
It appears that after 6 years the problem is still relevant in the industry.

Besides market research, app analytics are widely used in the industry, without developers adequately understanding the consequences.
Similar to the implications of third-party trackers, research shows the implications of using popular third-party analytics services.
For instance, the study of Chen \etal exposes the vulnerabilities of analytics services. 
They manipulated user-profiles constructed by such services to influence the ads shown to the users~\cite{Chen14}.
Analytics services also pose data privacy challenges. 
Privacy policies of analytics services are often difficult to read and demand too much time to comprehend. 
As a consequence, Bhardwaj \etal mention that developers are usually not aware of what or how the data is collected by analytics tools~\cite{Bhar15}.
Nevertheless, using app analytics for improving app usability has also been demonstrated by Ferre \etal
They extended Google Analytics to store specific low-level user interactions of interest to further test the usability~\cite{Ferr17}.
Another approach to understanding app usage patterns is suggested by Tongaonkar \etal who use in-app advertisements instead of app analytics~\cite{Tong13}.

End-user review analysis has also gained much traction in academia in recent years. 
Different approaches suggested by researchers aim to help app developers improve their app's functionality directly from end-user suggestions~\cite{Fu13, Guzm14,Pani15, Li18, Dalp19}.
We observed that user review analysis is practiced in the industry but on a limited scale. 
The interviewees rely on app stores such as Google Play, which provides automated user review summaries and sentiment analyses.
The reason we believe, as also mentioned by P4, could be the paid access to the academic publications, which underlines the need to publicize the work beyond the academic context \eg writing tweets, online blogs, articles \etc

In conclusion, all practitioners agreed that there is a lack of beta testers to gain confidence in the app they develop.
Nevertheless, they were all unaware of opinion mining techniques recently proposed in academic publications.
They similarly did not know the implications of using app analytics and usage trackers as mentioned earlier.  


\subsection{App type}
\label{subsec:app-type}

To fulfil end-user requirements, one does not always need a native mobile app, but hybrid or Progressive Web Apps\footnote{\url{https://developers.google.com/web/progressive-web-apps/}} (PWAs) can be a viable option too.

\subsubsection{Empirical observations}

\emph{Survey.}
There were no dedicated questions on the role of app stores in our survey.
The theme emerged from the survey responses and it was subsequently added to the interview instrument for further discussion with practitioners.

\emph{Interview.}
Interviewees had different opinions regarding hybrid apps.
According to P1, performance issues are actually not a concern for the end users as they are not so sophisticated that they can easily identify such details.
Additionally, platform-specific customizations generate in most scenarios up to 30\% additional code, which is acceptable for him and justifies their purpose.
The interviewee believes that discussions regarding performance of hybrid apps are ``usually held by technical people who do not reflect the true audience.''

PWAs have recently gained some traction, so we asked the interviewees their opinions on PWAs. 
They mention that PWAs are easy deployed and thus provide potential to kick-start useful discussions while gathering requirements.
Most of the interviewees agreed on the potential of PWAs and according to them industry understands their potential, but they also mentioned that it will take some time until they become mature.
They said that PWAs are especially helpful as they break the dependency on the app stores, and let developers release changes as quickly as possible, giving them advantages of native apps such as push notifications.

\subsubsection{Discussion}
Cost has been one of the main determining factors for both end users and developers for platform selection.
As we mentioned earlier, the cost of mobile devices affects the user base for a specific platform.
On the other hand, from the developers' perspective, the budget can affect the platform selection.
We identified three key themes from our observations:
\begin{itemize}
    \item hybrid apps and PWAs are seen as alternatives to native apps but with some reservations as to their quality;
    \item previous arguments regarding hybrid apps are not necessarily true due to technological advancements; and
    \item PWAs have the potential to replace native and hybrid apps in certain cases for few obvious advantages.
\end{itemize}
\noindent
RE practitioners must also familiarize themselves with hybrid frameworks as they can potentially replace native apps, and can be built at a lower price in terms of effort and number of required expert developers.
Several studies advocate cross-platform development due to extra effort and time to market caused by platform fragmentation~\cite{Ribe12, Chri11}.
The main complaint about hybrid apps has been their inferior performance~\cite{Corr12}.
Performance issues seem to be a bit overrated as they are far from obvious for average users.
It is also evident from the literature as Malavolta \etal from their survey of 11,917 apps from Google Play conclude that the perceived performance difference between native and hybrid apps is negligible~\cite{Mala15}.
Another major concern about hybrid apps is platform-specific customization~\cite{Corr12}, about which our interviewees disagreed.
Such observations differ from previous publications in this area; the reason, we believe, is merely due to technological advancements in the past few years~\cite{Wass10, Serr13}.

PWAs, on the other hand, are a recent phenomenon, and they are expected to occupy the gap between hybrid apps and plain web sites.
They even could potentially replace hybrid apps in certain cases where the app's purpose is just about displaying information.
PWAs are especially helpful when the native features such as camera access are not important for the app.
Additionally, they also simplify the app publishing workflow by breaking the dependency on app stores.
In response to a lack of academic involvement in the area of PWAs, the work of Bi{\o}rn-Hansen \etal provides a performance and feature comparison between cross-platform mobile and progressive web apps; notably they demonstrated that the performance of PWAs was far superior in terms of launch time and time from app icon tap to toolbar rendering~\cite{Bior17}.
In a similar comparison study, Cardieri \etal explored the aspects of user experience on three different platforms \ie native, web mobile and PWA, and they reported overall positive user experience despite there being a few interaction issues~\cite{Card18}.
Luntovskyy \etal also compare native, web-based and hybrid apps, and report that PWAs can represent an efficient alternative to native mobile apps due to several advantages such as a simplified installation process, and reduced data volume consumption~\cite{Lunt18}.

In conclusion, hybrid apps and PWAs have evolved in recent years into strong alternatives to native app development.
Most practitioners seemed to be aware of hybrid frameworks but they mostly had negative impressions about them, regardless of the existence of studies proving that performance differences are, in fact, negligible.
We observed that a practitioner had at least heard of PWAs but had never considered it an an option to mainstream mobile app development, despite certain advantages they offer such as breaking the dependency on app stores.
To fulfill end-user requirements, a native app may not always be an appropriate fit; alternatives such as hybrid apps may solve the problem more elegantly. 

\begin{table}[]
  \centering
  \caption{Main findings}
  \label{tab:main-findings}
  \footnotesize
  \begin{tabular}{ll}
  Factor                                                                       & Findings                                                                                                                                                                                                                                                                                                                                                                                                                                                            \\ \bottomrule
  \begin{tabular}[c]{@{}l@{}}Platform and\\ third-party libraries\end{tabular} & \begin{tabular}[c]{@{}l@{}}* Platform selection depends on several factors such as \\ cost, budget, market research and platform ecosystem\\ * End users, customers and developers can have different \\ concerns regarding platform selection\\ * Functional requirements are affected while selecting a \\ particular platform, and the opposite is true as well, \\ sometimes resulting in favouring hybrid apps\end{tabular}                                    \\ \hline
  \begin{tabular}[c]{@{}l@{}}App stores and\\ their policies\end{tabular}      & \begin{tabular}[c]{@{}l@{}}* Customers have no concerns regarding app stores, \\ but they are important for both end-users and developers \\ as they simplify the app distribution and discovery process\\ * Different app stores have different business models \\ and policies\\ * Policy changes by app stores are affecting app \\ development companies by causing unintentional \\ rework or in worse cases app denial\end{tabular}                           \\ \hline
  \begin{tabular}[c]{@{}l@{}}Non-functional\\ requirements\end{tabular}        & \begin{tabular}[c]{@{}l@{}}* Eliciting NFRs requires experience and expertise\\ * In the case of mobile apps, NFRs such as data privacy \\ can be unintentionally compromised by selecting a certain \\ third-party library\\ * Device capabilities are an important factor fora pp’s \\ performance and acceptance\end{tabular}                                                                                                                                    \\ \hline
  Learning about end users                                                     & \begin{tabular}[c]{@{}l@{}}* Market research is the client’s responsibility, \\ but it is rarely reliable\\ * Developers use other means to learn about the \\ user requirements such as through beta testing, \\ trackers, analytics, and app reviews\\ * Beta testing suffers from a lack of access to the \\ actual users; trackers and analytics have privacy breaches\\ * App review analysis is more popular in academia than \\ in the industry\end{tabular} \\ \hline
  App type                                                                     & \begin{tabular}[c]{@{}l@{}}* Hybrid apps and PWAs are seen as alternatives \\ to native apps but with some reservations as to \\ their quality\\ * Previous arguments regarding hybrid apps are \\ not necessarily true due to technological \\ advancements\\ * PWAs have the potential to replace native and \\ hybrid apps in certain cases for few obvious advantages\end{tabular}                                                                              \\ \bottomrule
  \end{tabular}
  \end{table}

%
%

\section{Related Work}
\label{sec:related-work}

Numerous papers outline differences between mobile and general-purpose software development~\cite{Fran17, Joor13}.

The study of Frances~\etal investigates aspects related to the development and management of mobile apps.
They interviewed four IT managers with experience in mobile app development, and surveyed 510 mobile app developers~\cite{Fran17}.
Only a small part of the study explores requirements issues regarding mobile apps, and unlike our study, it is limited to the discussion about usability aspects and subsequent GUI optimizations.
It explains app maintenance issues and reports that an additional business contract is required with the customer.
They present similar challenges with testing as we found in our work, \eg \emph{``beta testers are not available readily, and when at all they are in dozens and not in thousands,''} and \emph{``manual testing is prevalent.''}
Finally, they motivate the need for mobilization \ie using custom mobile apps to fulfil business needs of communication or data processing of organizational processes, which in our opinion is becoming more relevant.

Joorabchi~\etal  interviewed 12 senior mobile app developers and surveyed 188 mobile developers to gain an understanding of the challenges faced by developers~\cite{Joor13}.
Their findings, unlike ours, are mainly related to the challenges regarding app development for different platforms (at that time) and challenges regarding testing and analysis.
They argued that device fragmentation is not only a challenge for development, but also for testing.
Besides these aspects, they also pointed out app store requirements such as \emph{``changing app store policies''} play an important role in mobile app development.
We discuss this issue from the app's requirements point of view.

Nagappan~\etal shed light on the current and future research trends for various stages in the mobile app development life-cycle, \eg requirements, design and development, testing, and maintenance~\cite{Naga16}.
They extensively discuss the role of app stores in the mobile app industry, and also list few limitations such as \emph{``only a subset of user reviews available''}, and \emph{``no access to source code of the apps''}.
However, they neither explore these issues from the requirements point of view, nor do they discuss how developers tackle such issues.
This work also extensively discusses the battery usage of the apps, but interestingly none of the interviewees in our study has such concerns.

Wasserman \etal discuss software engineering issues specific to mobile app development, specifically citing performance, reliability, quality, and security among the most important NFRs for mobile apps.
However, their future direction regarding NFRs is limited to discovering differences between responsiveness, and data integrity in case of low battery, to name a few~\cite{Wass10}.
Dehlinger \etal outline and discuss four challenges that mobile application engineering faces, among which are the necessity of creating user interfaces for differently abled users, and platform fragmentation~\cite{Dehl11}.
As for the requirements, they signify the need for self-adapting apps based on the context \eg providing limited functionality for location-based services while having low battery rather than providing no service at all.

There are several studies that are relevant to this paper, but none of them exclusively target mobile development.
For instance, recent research has discussed the challenges with current RE practices~\cite{Groe17, Schon17, Naga16, Maal16, Inay15}.
Groen~\etal proposed a CrowdRE approach \ie performing requirements engineering with the crowd of stakeholders~\cite{Groe17}.
They mention that data privacy issues become prevalent while using different user feedback channels as the chance of exposing sensitive information increases.
Schon~\etal reported four relevant key challenges in their Agile RE approach: dependency issues arising between multiple teams due to coordination effort, customers not willing to let developers make independent decisions, inadequate access to the end users, and stakeholder participation throughout the development process~\cite{Schon17}.
Inayat~\etal explain additional issues such as budget and time estimation, negligence regarding NFRs, and fixed-priced contracts~\cite{Inay15}.
Maalej~\etal discuss data-driven RE, which involves user feedback analysis to identify potential requirements~\cite{Maal16}.
They report that managing huge user input requires substantial human resources for manual processing.
Moreover, scalability and the sophisticated tool support are also questionable.

There exist several relatively old studies which have outlined the practices and challenges with contemporary RE practices~\cite{Davi06, Zowg05, Hick03}.
The study from Davis~\etal questions the effectiveness of the RE techniques~\cite{Davi06}.
On the other hand, the study of Zowghi~\etal highlights the need to overcome the gap between research and industry, and also between novices and experts, as a potential challenge to effective RE~\cite{Zowg05}.
The empirical study from Hickey~\etal reports on when to use which elicitation technique~\cite{Hick03}.

%
%

\section{Threats to validity}
\label{sec:threats-to-validity}

The validity of the findings of a qualitative study is invariably subject to several threats~\cite{Wohl12}.

\emph{Construct validity} is threatened if the answers from the participants do not accurately reflect the real practice.
This could be due to the interviewees not feeling comfortable talking about certain topics or to the interviewer influencing the discussion.
To reduce these risks, we avoided judgment and evaluations during the discussions by assuring the interviewees anonymity in the study, and we abstained from communicating our assumptions to the participants.
In order to collect reliable data, we only selected participants who are knowledgeable about the companies' practices in gathering requirements.
When we interviewed multiple participants from the same company, we asked them not to talk about the interview to others in order to avoid any bias.
Unclear questions and misunderstandings between the interviewee and interviewers are also possible threats that cannot be completely ruled out.
We mitigated these threats by discussing the interview questions together with other experienced researchers and by conducting a pilot interview.
We also tried to explain the questions differently to the interviewees whenever we believed that the participant was suffering from misunderstandings.
Reliability threats that relate to researcher bias, however, cannot be completely ruled out, because the interviews and analyses were conducted by a single researcher.

\emph{External validity} issues are related to the inability to generalize the results of the study beyond the studied companies.
In order to achieve reasonable generalization during the interview sampling, we selected companies that showed different characteristics in terms of size, the domain of operation, and to some extent location.
However, our study suffers from selection bias as eight out of ten companies were from a single country \ie Switzerland.
Similarly, we could interview only ten practitioners due to several reasons such as contacted experts not being available for an interview, or not being interested.
Additionally, most of the interviewees were also survey participants, and hence they were already exposed to the topic which adds additional selection bias to our study.
Nevertheless, only few interviewees were from our personal contact list.

\emph{Internal validity} is threatened if a causal relationship between treatment and outcome is wrongly established.
A possible factor that could negatively impact the internal validity in our case is the interview change.
To reduce the effect, we evaluated the interview questions through an internal validation and a pilot interview and revised them before starting the real data collection.

%
%

\section{Conclusion}
\label{sec:conclusion}

We surveyed 45 companies and interviewed ten experts in the area of mobile app development to understand the challenges of requirements gathering.
Notwithstanding mobile and (traditional) software development being different in general, we observed that they are quite similar from the requirements-gathering perspective.

Nevertheless, there exist several factors that are more specific to the mobile domain.
For instance, RE practitioners need a deep understanding of platform ecosystems, including app store policies and available third-party libraries to make good decisions, \eg  platform or feature selection, that fulfil the functional requirements.
Also, dealing with non-functional requirements such as security and usability requires delicate expertise and experience as mobile devices offer customization both in the hardware and the operating system.

Furthermore, we realized that practitioners in our study are mostly unaware of several techniques for requirements elicitation such as the use of mobile apps and mobile devices, whereas they are extensively discussed in academia.
Similarly, opinion-mining techniques are prevalent in the state-of-the-art literature, but practitioners rather rely on user review summaries provided by app stores.
The reason we identified is that practitioners who are responsible for requirements collection have no direct technical experience with mobile app development, instead, they come from business or marketing background.

\section{Acknowledgments}
We gratefully acknowledge the financial support of the Swiss National Science Foundation for the project
``Agile Software Assistance'' (SNSF project No. 200020-181973, Feb. 1, 2019 - April 30, 2022).
We also thank CHOOSE, the Swiss Group for Original and Outside-the-box Software Engineering of the Swiss Informatics Society, for its financial contribution to the presentation of this paper.

\bibliographystyle{ACM-Reference-Format}
\bibliography{sample-base}

\end{document}